\documentclass{PoS}

\newcommand{\be}{\begin{equation}}
\newcommand{\ee}{\end{equation}}
\newcommand{\bea}{\begin{eqnarray}}
\newcommand{\eea}{\end{eqnarray}}
\newcommand{\mres}{m_{\rm res}}
\usepackage{subfigure}

\title{Nucleon Form Factors with 2+1 Flavors of Domain Wall Fermions and All-Mode-Averaging}

\ShortTitle{Nucleon form factors with 2+1 flavors of DWF and AMA}

\author{\speaker{Meifeng Lin}\thanks{Current affiliation: Computational Science Center, Brookhaven National Laboratory, Upton, NY 11973, USA} \\
        Argonne Leadership Computing Facility, Argonne National Laboratory, Argonne, IL 60439, USA\\
        RIKEN BNL Research Center, Brookhaven National Laboratory, Upton, NY 11973, USA \\
        E-mail: \email{mlin@bnl.gov}}

\abstract{We report recent progress in the calculations of the isovector nucleon electromagnetic form factors using 2+1 flavors of domain wall fermions at pion masses of 170 MeV and 250 MeV. The lattice size is fixed at $32^3\times64$ with a lattice cutoff scale of 1.37(1) GeV. For the calculations with $M_\pi = 170$ MeV, we employed the All-Mode-Averaging (AMA) technique, which led to roughly a factor of 20 improvement in computational efficiency and has reduced the statistical errors in our results significantly. We were also able to do calculations at two different source-sink separations, at roughly 1.3 fm and 1.0 fm, without much additional cost by reusing the low eigen-modes stored for the AMA calculations. We will present results for the isovector form factors and their derived quantities, including the Dirac and Pauli radii, anomalous magnetic moment and discuss the effects of possible excited-state contaminations. Connected contributions to the isoscalar Dirac and Pauli form factors will also be shown. 
}
\FullConference{31st International Symposium on Lattice Field Theory - LATTICE 2013\\
		July 29 - August 3, 2013\\
		Mainz, Germany}

\begin{document}

\section{Introduction}
Nucleon Dirac and Pauli form factors, $F_1^N(Q^2)$ and $F_2^N(Q^2)$ respectively, are defined through the nucleon vector matrix elements,
\begin{eqnarray}
\langle N(p') | J_\mu^N(x) | N(p) \rangle = e^{i(p'-p)\cdot x }\overline(u)(p') \left [ \gamma_\mu F_1^N(Q^2) + i \sigma_{\mu\nu} \frac{q_\nu}{2M_N} F_2^N(Q^2)\right ] u(p), \\
N = \mathrm{proton} (p)\,\,\,\, \mathrm{or}\,\,\,\, \mathrm{neutron}(n)\nonumber,
\end{eqnarray}
where $p$ and $p'$ are the initial and final momenta of the nucleon, respectively. $Q^2=-(p'-p)^2$ is the momentum transfer from the incoming nucleon to the outgoing nucleon. The Dirac and Pauli form factors are related to the electric and magnetic Sachs form factors by $G^N_E(Q^2) = F^N_1(Q^2) - \frac{Q^2}{4M_N^2} F^N_2(Q^2)$, and $G^N_M(Q^2) = F^N_1(Q^2) + F^N_2(Q^2)$. 

At $Q^2=0$, the Dirac and Pauli form factors encode the charge and anomalous magnetic moment of the nucleon. That is, $e_N = F_1^N(0)$ and $\kappa_N = F_2^N(0)$. Experimentally, we know $e_N = 1$ in units of electron charge. On the lattice, we use this fact to determine the vector renormalization constant $Z_V$. The anomalous magnetic moment $\kappa_N$ is also known precisely experimentally~\cite{Beringer:1900zz}, with $\kappa_p = 2.792847356 (23)$ for the proton, and $\kappa_n = -1.913042(5)$ for the neutron. The slopes of $F_1^N(Q^2)$ and $F_2^N(Q^2)$ at $Q^2=0$ give the mean squared charge and magnetic radii of the nucleon,
\begin{equation}
\langle( r^N_i)^2\rangle = -6 \frac{d F^N_i(q^2)}{dQ^2} |_{Q^2=0}, \,\,\, i = 1,2. 
\end{equation}
The calculation of $\langle (r^p_1)^2 \rangle$ is particularly interesting, as it is related to the proton electric charge radius $\langle r^p_E \rangle$ through 
$\langle (r_1^p)^2\rangle = \langle (r_E^p)^2 \rangle - \frac{6}{4M_p^2} \kappa_p$. The CODATA value for $r_E^p \equiv \langle (r_E^N)^2 \rangle^{1/2}$ from electron-proton scattering experiments is 0.8775(51) fm~\cite{Bernauer:2010wm}. But a newer type of experiment using the Lamb shift of muonic hydrogen gave a value 0.84087(39) fm \cite{Antognini:1900ns} which is about seven standard deviations from the CODATA result. This discrepancy has been widely known as \emph{the proton size puzzle} and has spurred a lot of discussions in both the experiment and theory communities. 

Lattice QCD calculation is the only non-perturbative method to compute nucleon form factors and their derived quantities with controllable systematic errors. In the context of the proton size puzzle, lattice results, when sufficient precision is reached, can provide valuable input from QCD predictions to help resolve the discrepancy. However, before we can make contact with the experiments, we have to address various systematic errors associated with lattice calculations, such as the chiral extrapolation, finite volume, discretization errors and possible excited-state contaminations. We attempt to minimize sources of systematic errors by performing the calculations at light pion masses in a volume of (4.6 fm)$^3$ and studying the possible excited-state effects with two different source-sink separations. 

While the techniques for the lattice calculation of nucleon form factors have been well established in the past decade, computational challenges still remain for direct calculations at the physical quark mass. Over the past few years, 2+1-flavor lattice QCD simulations have increasingly been performed at light pion masses very close to the physical point, thanks to the improvements in numerical algorithms and computing power. Here we report one such calculation at pion masses as light as 170 MeV with a new noise-reduction technique, All-Mode-Averaging (AMA)~\cite{Blum:2012uh}. The results for the nucleon axial charge and bare structure functions from the same calculation are shown in \cite{Ohta:2013qda}. In this report we will present the results for the nucleon vector form factors.

The calculation was performed with 2+1 flavors of domain wall fermions (DWF) in a volume of $32^3\times64$ with $L_s = 32$. The gauge configurations~\cite{Arthur:2012opa} were generated with the Iwasaki gauge action with the Dislocation Suppressing Determinant Ratio (DSDR) at $\beta = 1.75$, giving rise to a lattice cutoff of $a^{-1}=1.37(1)$ GeV, or $a \approx 0.144$ fm. The residual mass with these parameters is determined to be $a m_{\rm res}=0.001842(7)$. The dynamical pion masses are about 170 MeV and 250 MeV with the input light quark masses of $am_l = 0.001$ and $0.0042$ respectively.

\section{Noise Reduction with All-Mode-Averaging}
Lattice calculations with nucleon states are particularly challenging numerically, because the signal-to-noise decreases exponentially with a decreasing pion mass due to the three-pion contributions to the noise. N\"aively, the signal-to-noise, $S/N$, for the nucleon two-point correlation function follows
\begin{equation}
S/N (t) \propto \sqrt{N_{\rm meas}} \exp\left [ -(M_N - \frac{3}{2} M_\pi) t \right ], 
\end{equation}
where $N_{\rm meas.}$ is the number of independent measurements. The All-Mode-Averaging technique~\cite{Blum:2012uh} allows us to increase $N_{\rm meas}$ significantly without adding much more cost. This is possible by constructing an improved operator $O^{\rm imp}$ which consists of a cheaply calculated, but less precise, approximate operator $O^{\rm apprx}$ and a correction term $O^{\rm rest}$ that compensates for the bias that may be introduced in $O^{\rm apprx}$: $O^{\rm imp} = O^{\rm rest} + O^{\rm apprx}$.

As $O^{\rm apprx}$ is much cheaper to calculate by design, we can perform many calculations of $O^{\rm apprx}$, and rely on less frequently calculated, more precise, original operator $O^{\rm exact}$ to compute the correction
\be
O^{\rm rest} = O^{\rm exact} - O^{\rm apprx}. 
\ee
As long as $O^{\rm apprx}$ has strong correlations with $O^{\rm exact}$, the correction term $O^{\rm rest}$ should be small, and the statistical noise will largely be determined by the number of measurements for $O^{\rm apprx}$.

In the implementation for the nucleon calculations presented here, we first computed 1000 low-lying eigenmodes of the DWF Dirac operator, and then use these low eigenmodes to compute the low-mode part of the propagator, $S_l^\parallel$. The low-mode-deflated quark propagator $S_{\rm sloppy}^{\perp}$ is then computed to a \emph{sloppy} stopping condition of order $10^{-3}$, giving an approximate quark propagator $S_{\rm apprx} = S_l^{\parallel} + S_{\rm sloppy}^{\perp}$, from which the approximate nucleon correlation functions are constructed. 

The approximate calculations were done on 7 time slices and 16 spatial source locations on each time slice, giving a total of 112 measurements per configuration. The exact calculations were done on 4 time slices, with one spatial source location per time slice. The all-mode-averaged (AMA) result for the observable $O$ is then
\be
O_{\rm AMA} = \frac{1}{N_{\rm apprx} }\sum_{i=1}^{N_{\rm apprx}} O_{\rm apprx}^i + \frac{1}{N_{\rm exact}} \sum_{j=1}^{N_{\rm exact}} \left ( O_{\rm exact}^j - O_{\rm apprx}^j \right ). 
\ee
To further reduce the cost, we use the M\"obius domain wall fermion operator in $O_{\rm apprx}$ with a smaller $L_s=16$. The parameters of the M\"obius DWF operator were chosen such that the valence residual mass roughly matches that in the dynamical simulation with the standard Shamir DWF operator. 

\begin{figure}[htbp]
\centering
\subfigure[Plateau of $F_1^{p-n}$ at the first non-zero momentum with $M_\pi=170$ MeV. Points are shifted slightly for clarity.]{
  \includegraphics[width=0.45\textwidth]{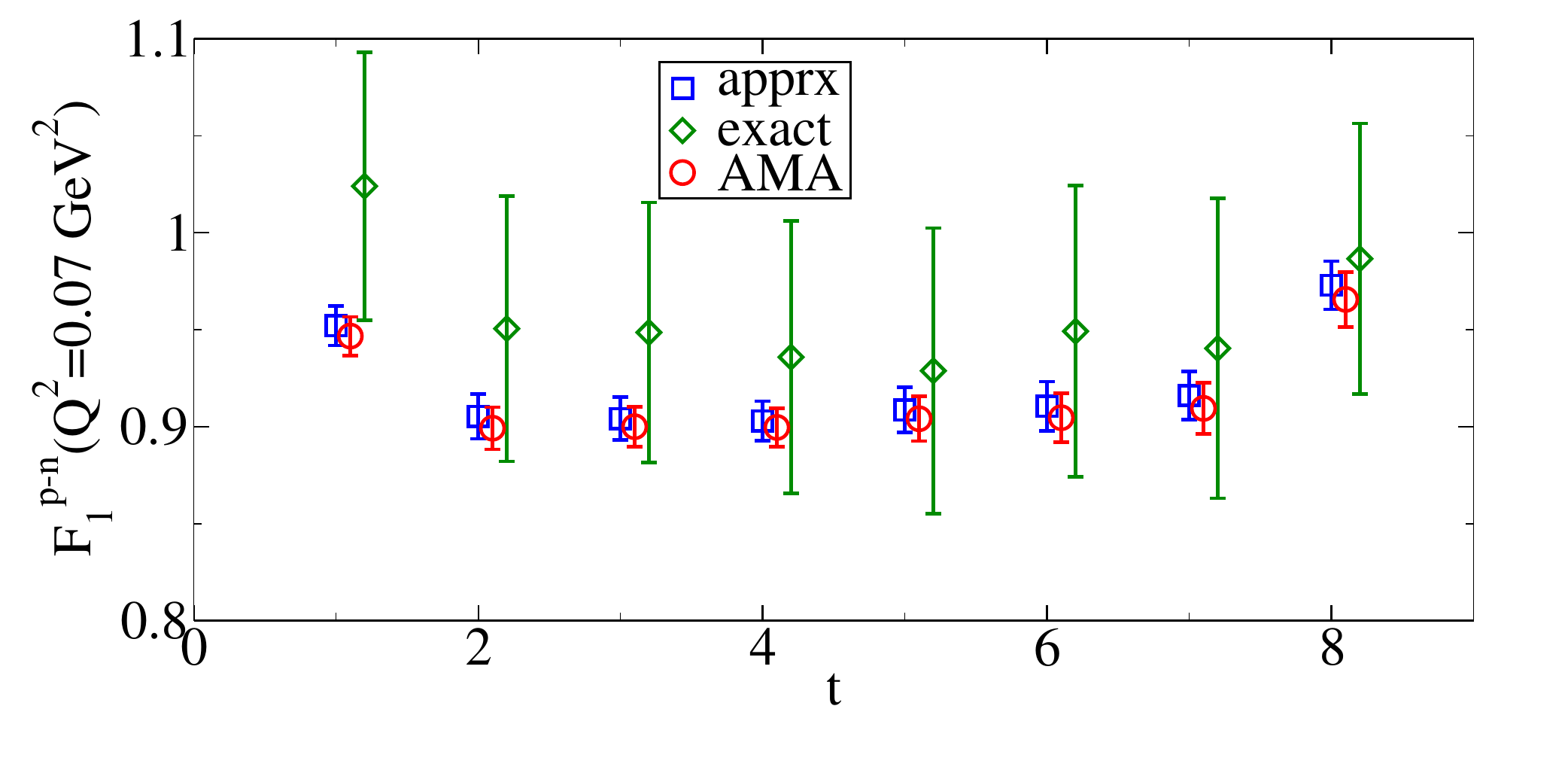}
   \label{fig:AMAcompF1}
   }
\hfill
\subfigure[Isovector Dirac form factor, $F_1^{p-n}(Q^2)$, with $M_\pi = 170$ MeV. Points are shifted slightly for clarity.]{
  \includegraphics[width=0.45\textwidth]{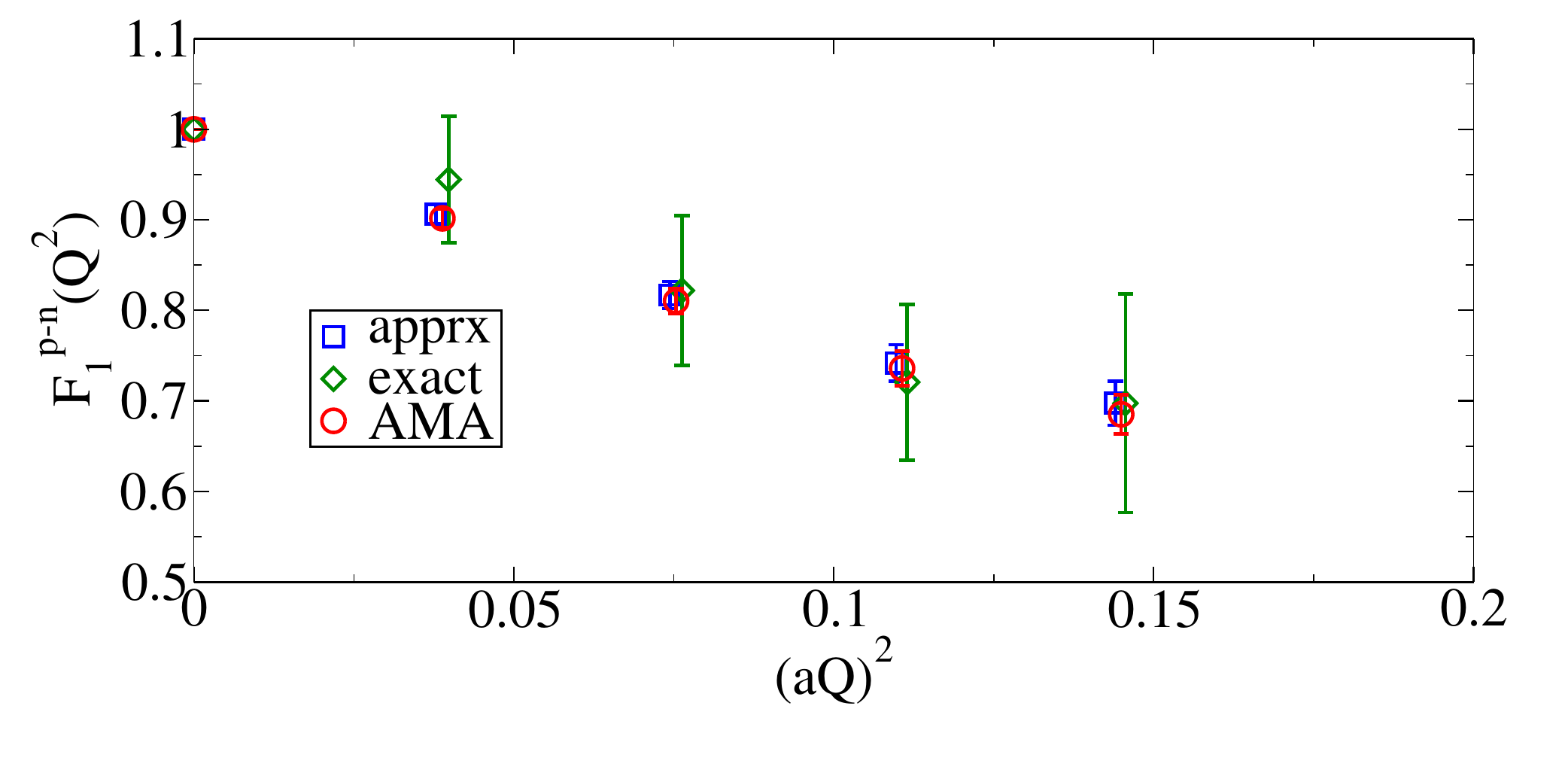}
   \label{fig:AMAcompF1Q}
   }
\caption{Comparison of the exact, approximate and all-mode-averaged results on the light ensemble with a pion mass of about 170 MeV.}
\label{fig:AMAcomp}
\end{figure}

In Figure~\ref{fig:AMAcomp} we show the comparison of the exact, approximate and all-mode-averaged results for the plateau of $F_1^{p-n}$ at the first non-zero momentum (Figure~\ref{fig:AMAcompF1})  and for the Dirac form factors over the whole $Q^2$ range. The statistical errors are reduced by a factor of 4.6 to 6.4 over the whole $Q^2$ range (Figure~\ref{fig:AMAcompF1Q}). N\"aively this would require a factor of 21 to 41 more computations if no improvements were implemented. As one sloppy calculation costs roughly about 1/65 of one exact calculation, taking into account the cost of the eigenmodes, the actual AMA cost is only 1.4 times that of the exact calculation without deflation. In this example, the speedup with AMA is 15 to 29 times.

\section{Preliminary Results}
We have increased the statistics for the 250 MeV ensemble from four source locations per lattice to seven source locations per lattice ($t_{\rm src} = 8 n, n = 0, ..., 6$). While we did not employ AMA in this ensemble, we sped up the calculations with the coherent-sink sequential propagators~\cite{Bratt:2008uf,Bratt:2010jn} for the source locations at $t_{\rm src} = 8,  24$ and $40$. For the calculations with $M_\pi=170$ MeV, we performed 4 exact calculations and 112 sloppy calculations per lattice on 39 configurations. The details of the calculation are summarized in Table~\ref{tab:params}. For the error analysis, we blocked different sources on each lattice, and treated different lattice configurations as independent. Further blocking consecutive configurations did not increase the statistical errors significantly. 

 \begin{table}[ht]
        \centering
        \begin{tabular}{ccccccccc}
    
          \hline
          \hline
          $am_l$ & $am_s$ & $L^3\times T$ & $L_s$ & {$m_\pi$ [MeV]} & {$m_\pi L$} & $a\mres$ & \# of configs. & \# of meas.\\
          \hline
          0.001 & 0.045 & $32^3\times64$ & 32 & {170} & {4.0} & 0.0018 & 39 & 4368\thanks{The number indicates the total number of measurements with the sloppy CG. The number of exact measurements is 4 per configuration, or 156.} \\
          \hline
          0.0042 & 0.045 & $32^3\times 64$ & 32 & {250} &{5.8} & 0.0018 & 165 & 1155 \\
           \hline
          \hline
        \end{tabular}
        \caption{Details of the calculations.}
        \label{tab:params}
      \end{table} 

Our results for the isovector Dirac and Pauli form factors can be fit with the empirical dipole form: $F_i = A_i/ (1+Q^2/M_i^2)^2$, from which we obtain the results for the isovector Dirac radius $\langle r_1^2\rangle^{1/2}_{p-n}$, Pauli radius $\langle r_2^2 \rangle^{1/2}_{p-n}$ and the anomalous magnetic moment $\kappa_{p-n}$. These results, together with the previous calculations with 2-flavor~\cite{Lin:2008uz} and 2+1-flavor~\cite{Yamazaki:2009zq} domain wall fermions, are shown in Figure~\ref{fig:results}. We also show the comparison between the 2012 results~\cite{Lin:2012jz} without AMA (brown empty diamonds) and this year's improved results (red filled diamonds). It is clear that AMA has offered substantial error reduction in the calculations with $M_\pi=170$ MeV. While the results for the isovector Pauli radius and the anomalous magnetic moment are within two standard deviations of the experimental values at $M_\pi=170$ MeV, the isovector Dirac radius still shows a 20\% deficit. In Ref.\cite{Green:2012ud} the authors found that excited-state contaminations tend to result in a small value for $\langle r_1^2\rangle^{1/2}_{p-n}$. As we will discuss in Section \ref{sec:excited-state}, our calculations do not seem to suffer from excited-state contaminations. The deficit here may be due to a large chiral log near the physical pion mass. 

\begin{figure}[ht]
\centering
\subfigure[]{
  \includegraphics[width=0.31\textwidth]{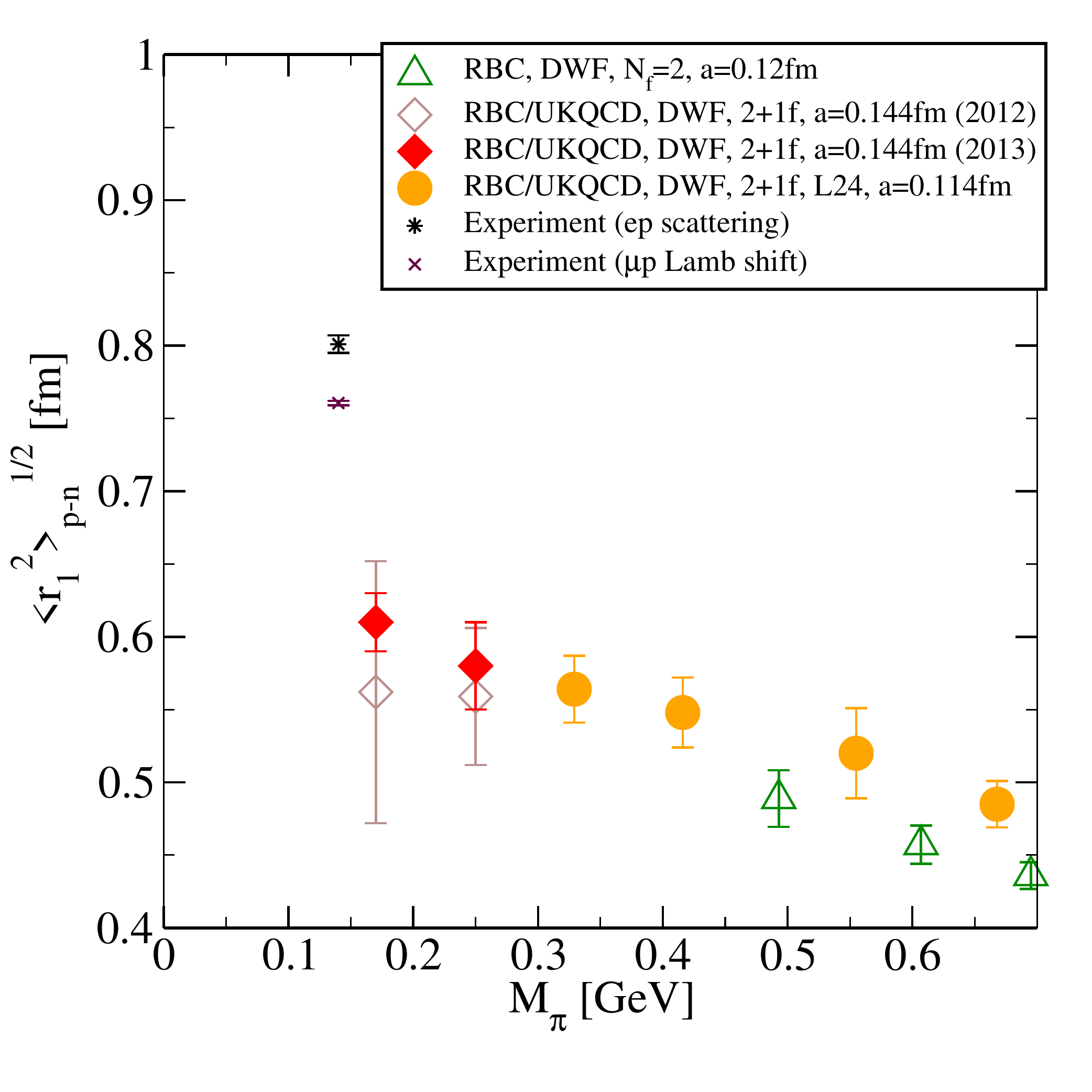}
   \label{fig:r1}
   }
\hfill
\subfigure[]{
  \includegraphics[width=0.31\textwidth]{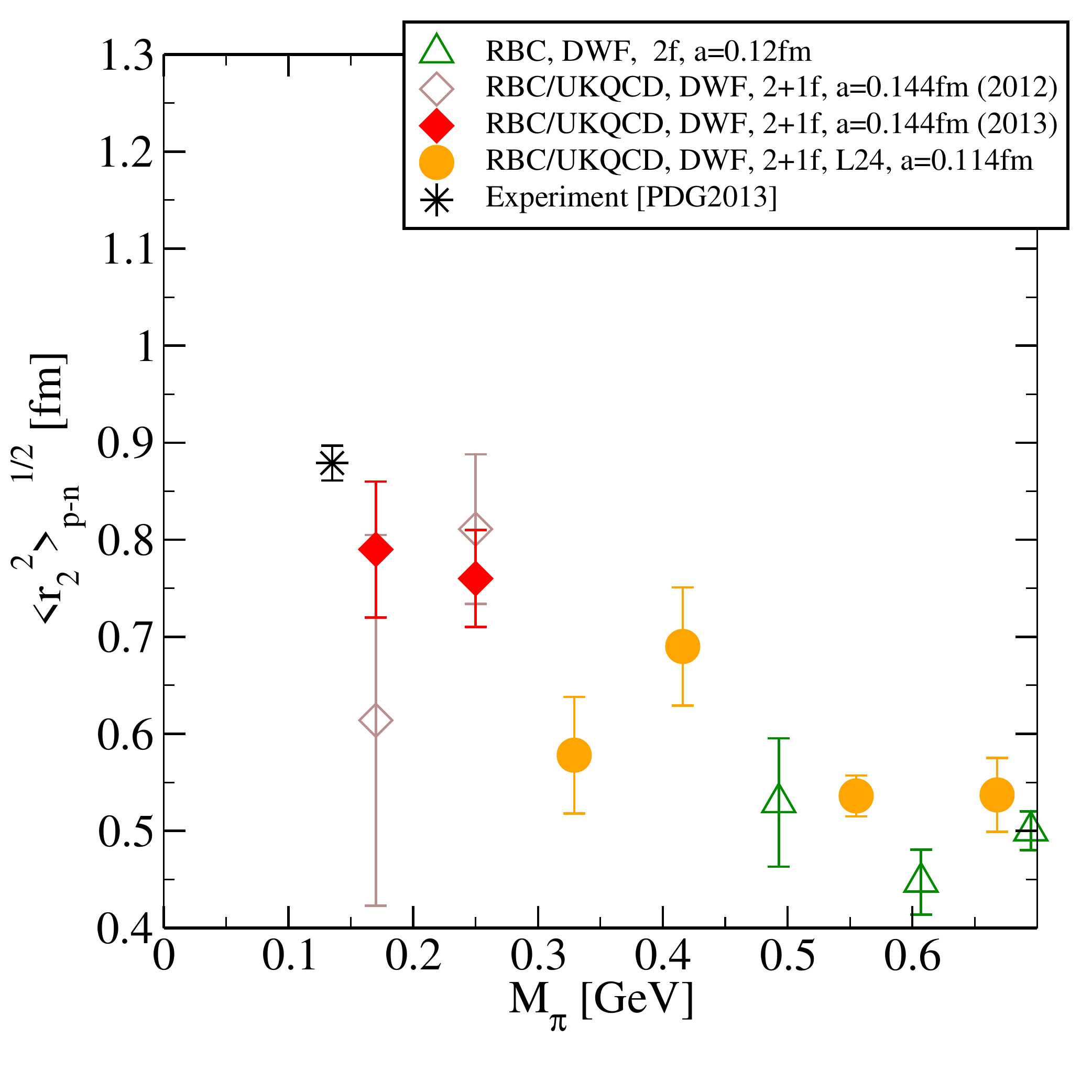}
   \label{fig:r2}
   }
\hfill
\subfigure[]{
  \includegraphics[width=0.31\textwidth]{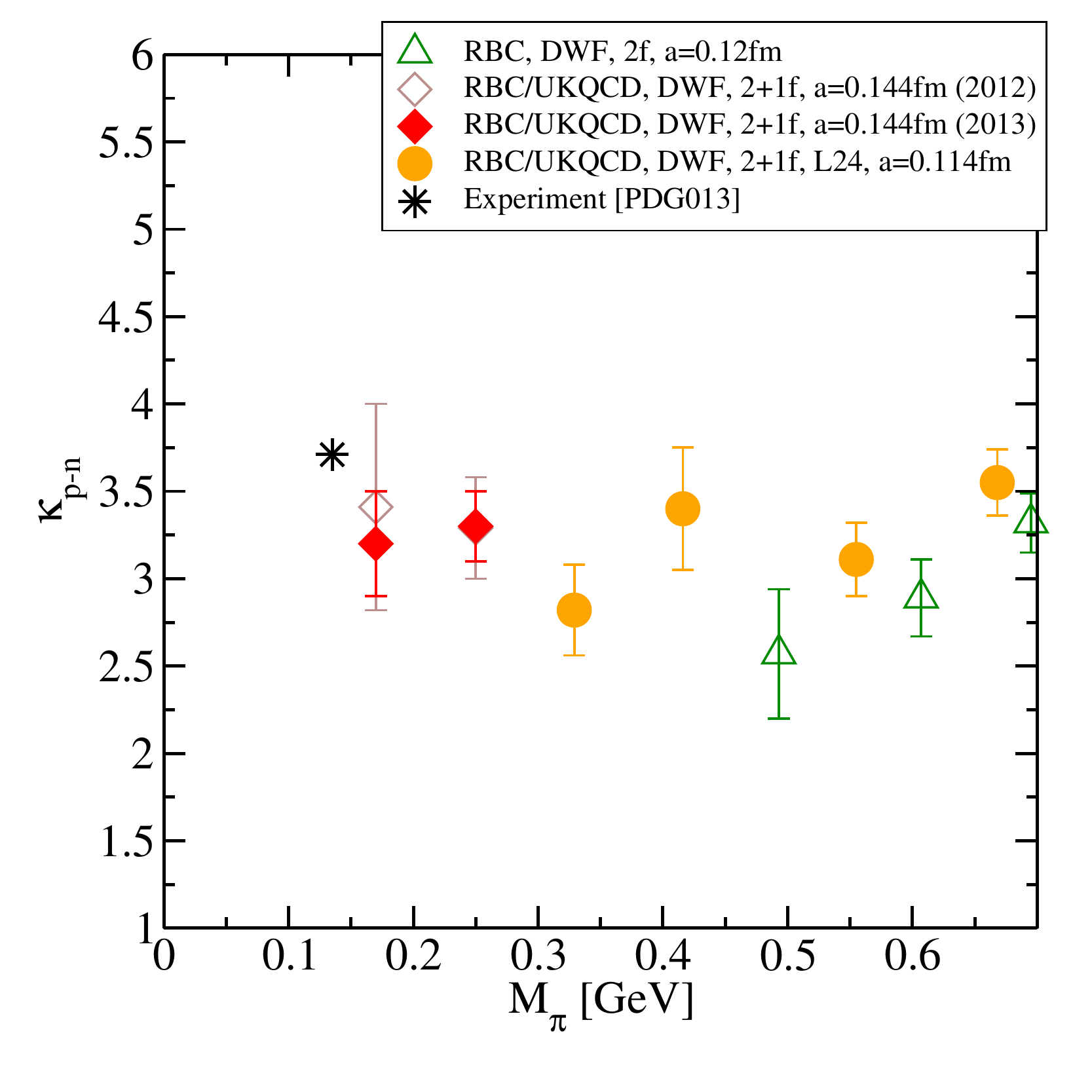}
   \label{fig:kappa}
   }
\caption{Preliminary results for isovector nucleon Dirac radius (left), Pauli radius (center) and anomalous magnetic moment (right).}
\label{fig:results}
\end{figure}

\begin{figure}[ht]
\centering
\subfigure[]{
  \includegraphics[width=0.31\textwidth]{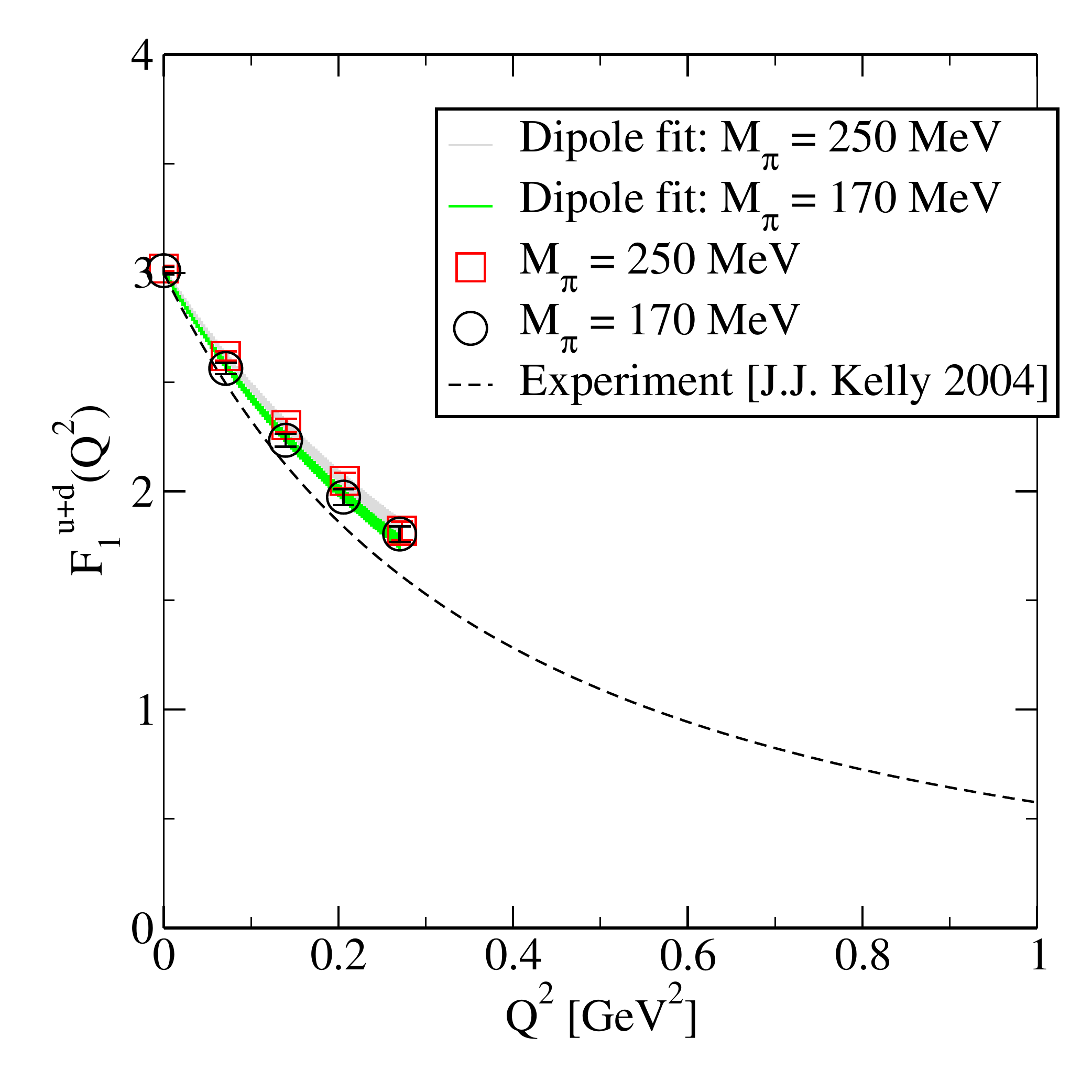}
   \label{fig:F1upd}
   }
\hfill
\subfigure[]{
  \includegraphics[width=0.31\textwidth]{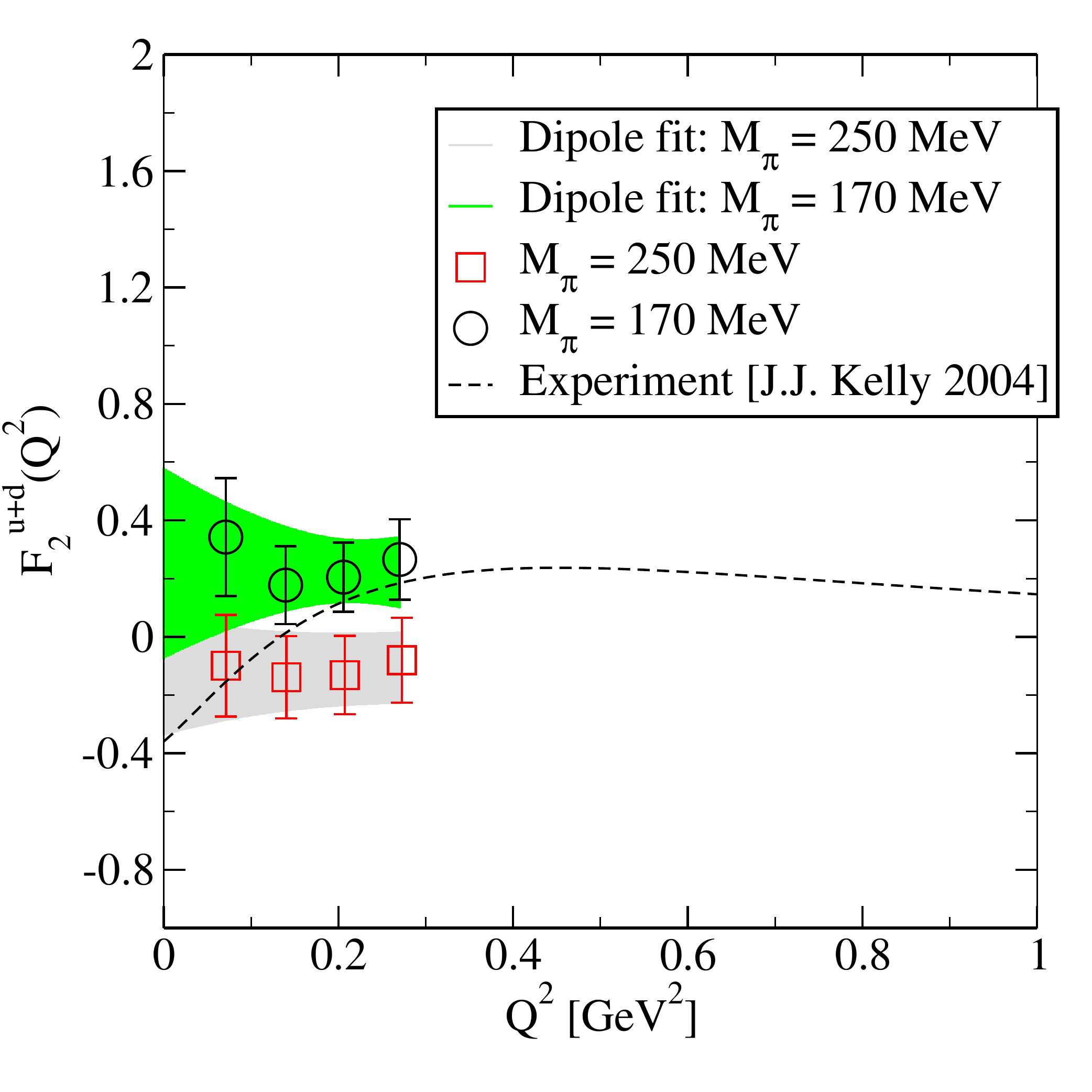}
   \label{fig:F2upd}
   }
\hfill
\subfigure[]{
  \includegraphics[width=0.31\textwidth]{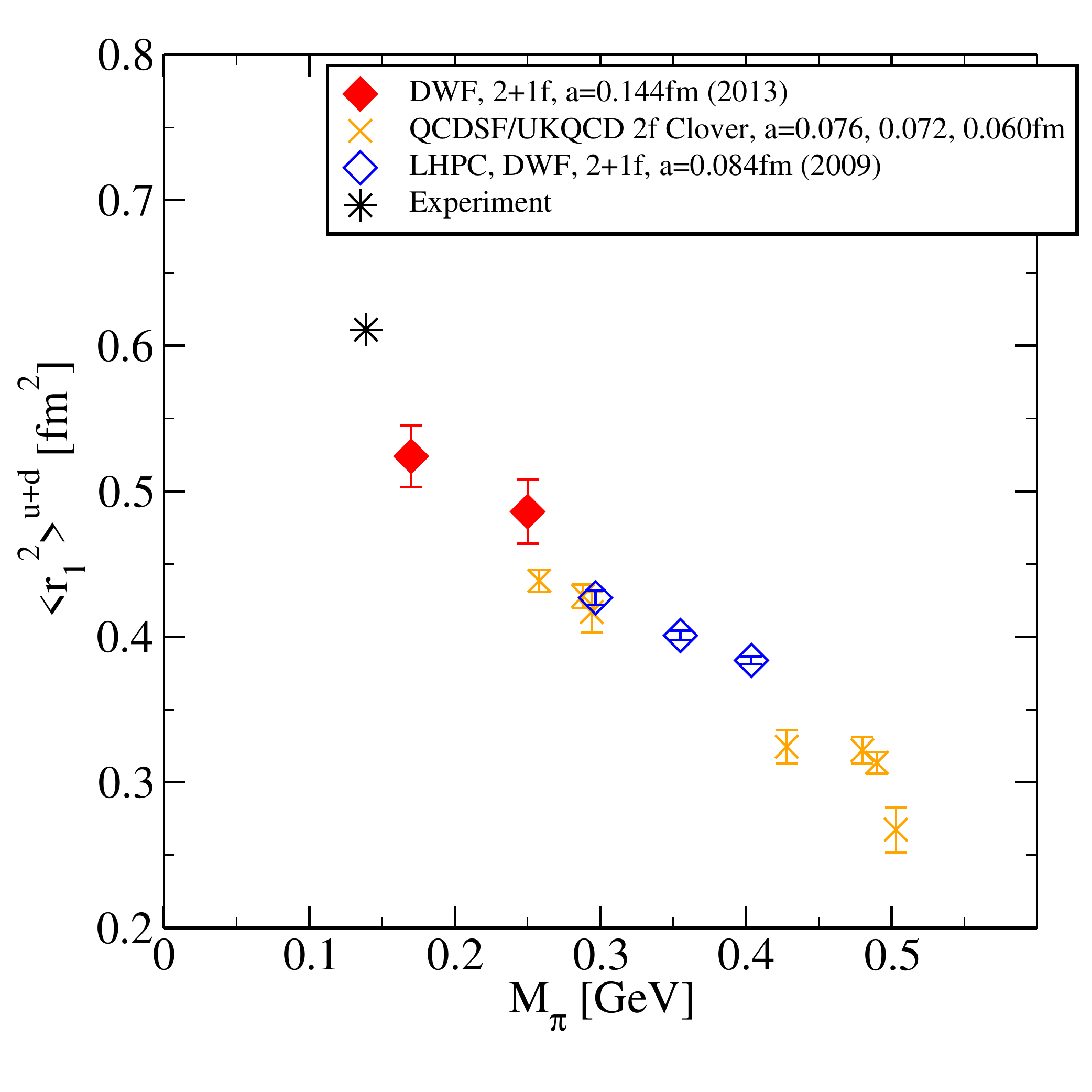}
   \label{fig:r1s2}
   }
\caption{Connected contribution to the isoscalar nucleon Dirac form factors (left), Pauli form factors (center) and the Dirac radius (right).The dashed lines are the parameterizations of the experimental data in Ref.\cite{Kelly:2004hm}.}
\label{fig:results}
\end{figure}

While we did not include disconnected diagrams in our calculations, we can still look at the contribution of the connected diagrams to the isoscalar form factors  $F_1^{u+d}(Q^2)$ and  $F_2^{u+d}(Q^2)$, shown in Figure \ref{fig:F1upd} and Figure \ref{fig:F2upd} respectively. The shaded curves are from dipole fits. While $F_1^{u+d}(Q^2)$ fits well to the dipole form and allows us to determine the isoscalar Dirac radius quite well, $F_2^{u+d}(Q^2)$ shows little curvature and the extracted values for the radius and anomalous magnetic moment are consistent with zero. In Figure \ref{fig:r1s2} we show our results for the isoscalar Dirac radius, together with two other lattice calculations~\cite{Syritsyn:2009mx,Collins:2011mk}. As opposed to the isovector case, here our result for the isoscalar Dirac radius at $M_\pi = 170$ MeV approaches the experiment steeply.

\section{Study of Excited-State Contaminations}
\label{sec:excited-state}
\begin{figure}[ht]
\centering
\subfigure[Comparison of plateaus.]{
  \includegraphics[width=0.31\textwidth]{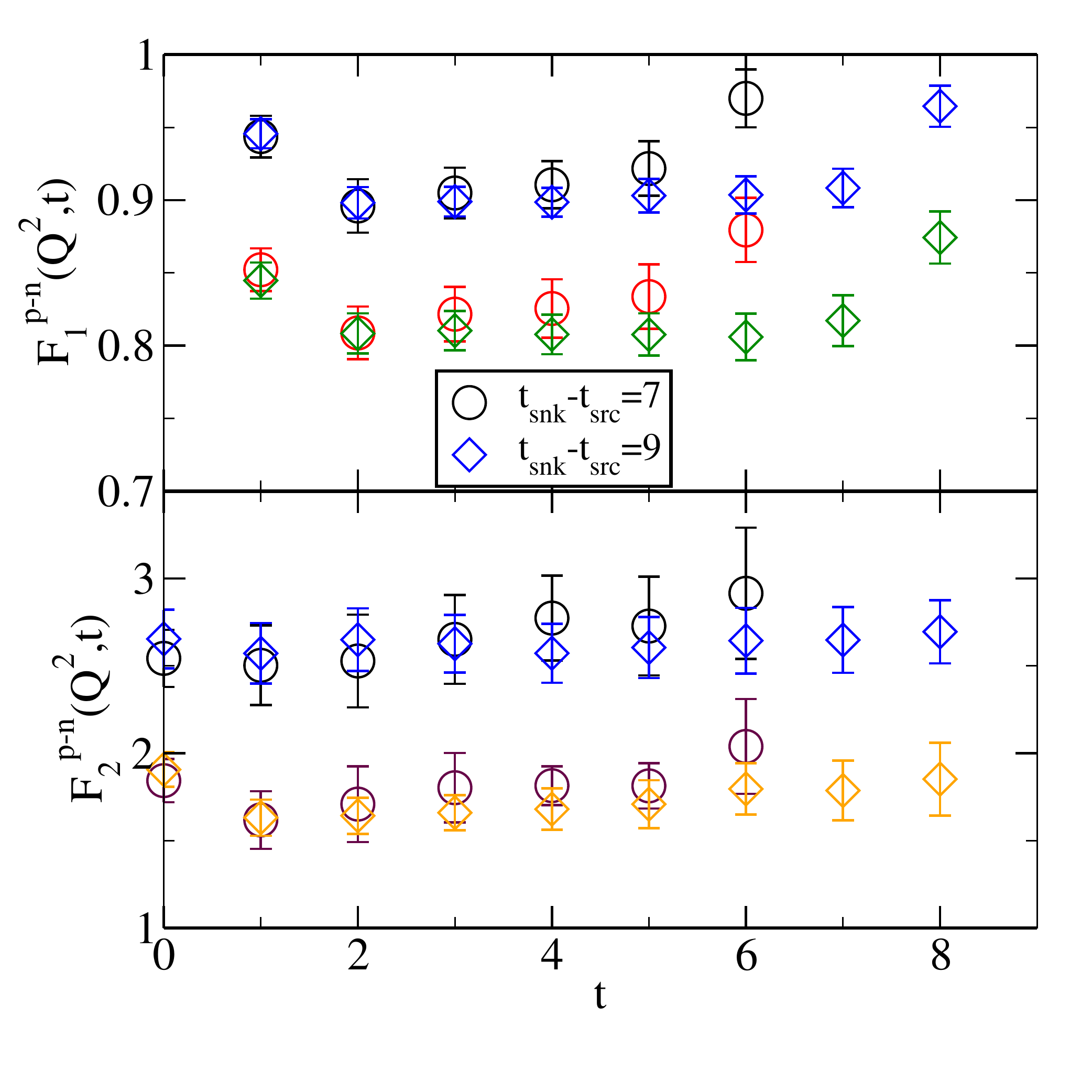}
   \label{fig:source-sink-plateau}
   }
\hfill
\subfigure[Comparison of $F_1^{p-n}(Q^2)$.]{
  \includegraphics[width=0.31\textwidth]{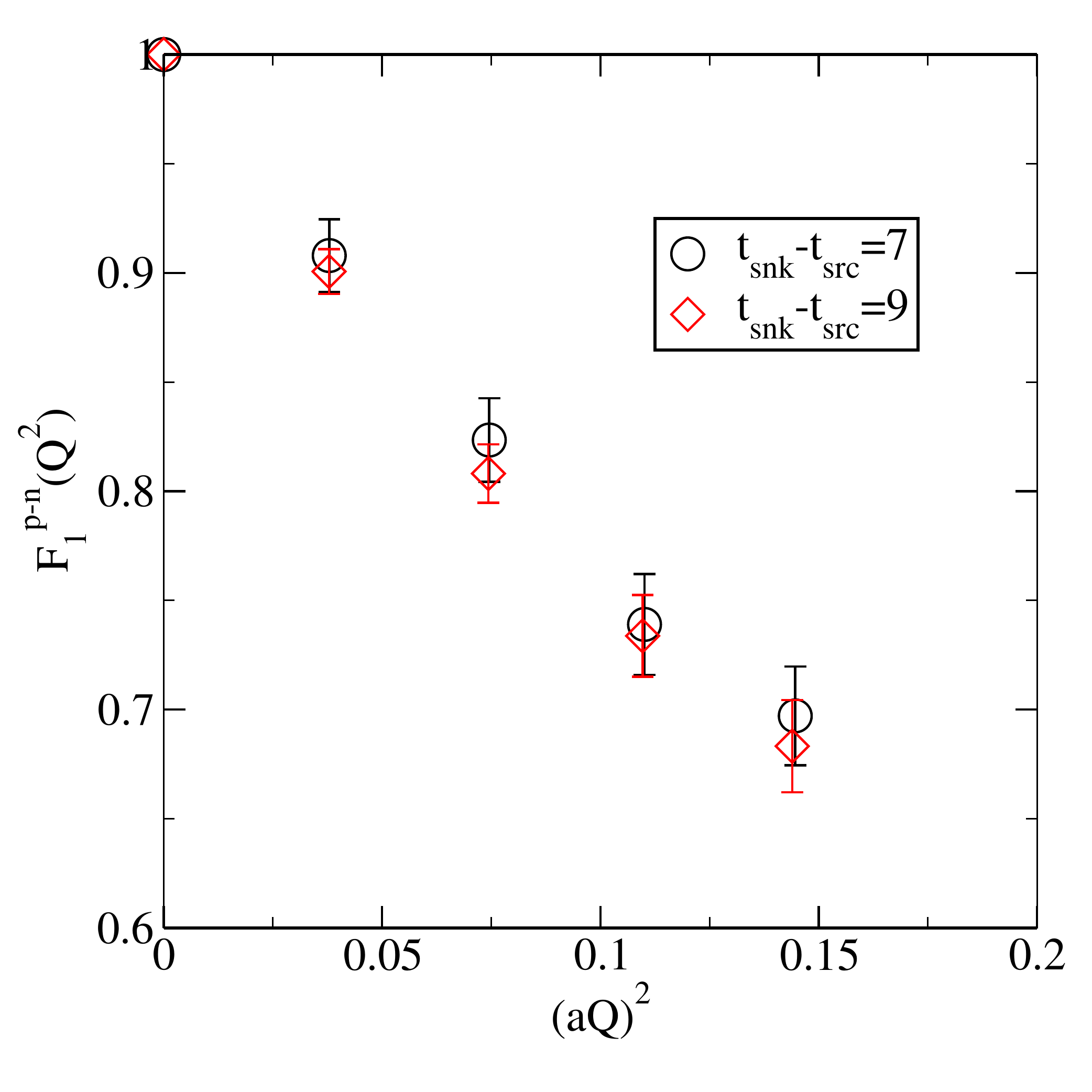}
   \label{fig:source-sink-F1}
   }
\hfill
\subfigure[Comparison of $F_2^{p-n}(Q^2)$.]{
  \includegraphics[width=0.31\textwidth]{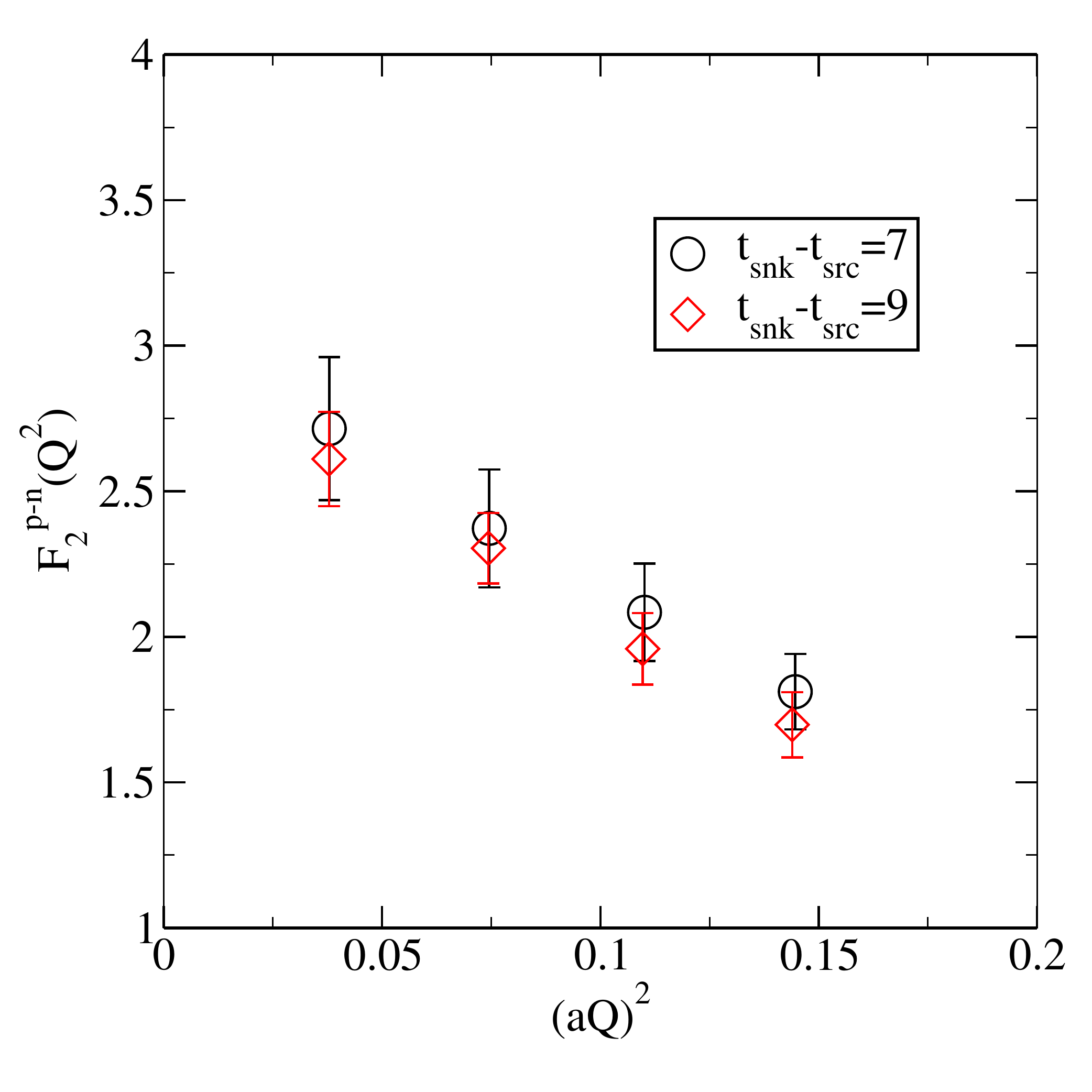}
   \label{fig:source-sink-F2}
   }
\caption{Comparisons of results with two different source-sink separations for the calculations at $M_\pi=170$ MeV.}
\label{fig:source-sink}
\end{figure}

AMA also allows us to generate nucleon three-point functions with different source-sink separations without much additional cost, since we can reuse the eigenmodes that we calculated at the beginning. We studied the possible excited-state contaminations on the light ensemble with $M_\pi = 170$ MeV by comparing the plateaus of the nucleon Dirac and Pauli form factors at the source-sink separations of 7 and 9 lattice units, corresponding to roughly 1 fm and 1.3 fm physical separations, respectively. For $t_{\rm snk}-t_{\rm src}=7$, eight configurations were used with 32 measurements per configuration. The comparisons of the plateaus for $F_1^{p-n}(Q^2, t)$ and $F_2^{p-n}(Q^2,t)$ at two representative $Q^2$ values in each case are shown in Figure~\ref{fig:source-sink-plateau}, from which we see no apparent excited-state contaminations. Fitting from $t=3-6$ for $t_{\rm snk}-t_{\rm src}=9$ and $t=3-4$ for $t_{\rm snk}-t_{\rm src}=7$, we obtain the results for $F_1^{p-n}(Q^2)$ and $F_2^{p-n}(Q^2)$ as shown in Figures \ref{fig:source-sink-F1} and \ref{fig:source-sink-F2}. The lack of exicted-state contaminations in our calculations may be attributed to the tuning of our nucleon source operator~\cite{Lin:2012jz}, which perhaps has a very good overlap with the nucleon ground state. It is possible that with increased statistics, the two source-sink separations may show statistically different results. But with the statistics available to us, we cannot identify any excited-state contaminations. Similar studies for the nucleon axial charge are presented in \cite{Ohta:2013qda}.

\section{Summary and Outlook}
We have shown that All-Mode-Averaging offers substantial speedup for nucleon structure calculations, which is about a factor of 20 in our application to the calculations with $M_\pi=170$ MeV. Our results for the Dirac and Pauli form factors have dramatically reduced statistical errors. Despite the light pion masses, our results for the isovector Dirac radius still show roughly a 20\% deficit compared to the experimental results. While the isovector Pauli radius and anomalous magnetic moment still suffer from large statistical errors, their results are within two standard deviations of their corresponding experimental values. We study the possible excited-state contaminations with two source-sink separations at $M_\pi=170$ MeV, and find no statistically significant effects. We also determine connected contributions to the isoscalar form factors, and find that the isoscalar Dirac radius (connected) shows a nice sharp rise towards the physical value. However, we cannot get a signal for the isoscalar Pauli radius or the anomalous magnetic moment. To address the deficit of the isovector Dirac radius, we are now in the process of doing the calculations directly at the physical pion mass, using 2+1-flavor DWF gauge configurations \cite{Blum:2013phy} generated by the RBC and UKQCD Collaborations.

\section*{Acknowledgments}
M.L. thanks Yasumichi Aoki, Tom Blum, Taku Izubuchi, Chulwoo Jung, Shigemi Ohta, , Shoichi Sasaki, Eigo Shintani and Takeshi Yamazaki for valuable contributions and discussions. The gauge configurations used in these calculations were generated by the RBC and UKQCD Collaborations on the Blue Gene/P supercomputers at Argonne Leadership Computing Facility, Brookhaven National Laboratory, University of Edinburgh and RIKEN-BNL Research Center. The majority of the quark propagators and correlation functions were calculated on computing clusters provided by US National Science Foundation through the XSEDE program and by the RIKEN Integrated Cluster of Clusters in Japan. M.L. was partially supported by SciDAC-3 and Argonne Leadership Computing Facility at Argonne National Laboratory under contract DE-AC02-06CH11357.

\end{document}